\documentclass[final,5p,times,twocolumn]{elsarticle}

\usepackage{subfigure}
\usepackage{graphicx}
\usepackage{epstopdf}
\usepackage{longtable}
\usepackage{amsmath}
\usepackage[load-configurations = abbreviations]{siunitx}
\usepackage{color}
\usepackage[usenames,dvipsnames,svgnames,table]{xcolor}
\usepackage{lineno,hyperref}
\usepackage{siunitx}

\modulolinenumbers[5]

\begin{document}

\author[LNF]{A. Marocchino \corref{mycorrespondingauthor} }
\cortext[correspondingauthor]{Corresponding author}
\ead{alberto.marocchino@lnf.infn.it,alberto.marocchino@gmail.com}
\author[LLR]{A. Beck}

\address[LNF]{Laboratori Nazionali di Frascati, INFN, Via E. Fermi 40, Frascati, Italy}
\address[LLR]{Laboratoire Leprince-Ringuet, \'{E}cole polytechnique, 91128 Palaiseau, France}

\title{Summary and report Working Group 6: theory and simulation, for the third edition of the European Advanced Accelerator Concept Conference}

\date{\today}

\begin{abstract}
The theory and simulation working group (working group 6) of the third edition of the European Advanced Accelerator Workshop has been characterized by a strong numerical connotation. Particle in cell codes have proven to be a necessary tool to finely investigate the underlying physics both for laser and plasma wakefield acceleration processes. This year, the section has been characterised by an interest in the limitation of the numerical Cherenkov effect, the mitigation of the hose-instability, start-to-end simulations and more generally on new numerical schemes and diagnostics.
\end{abstract}

\maketitle

\section{Introduction}
Working Group 6 (WG6), theory and simulation, of the third edition of the European Advanced Accelerator Workshop has been characterized by a strong numerical connotation: massively parallel particle in cell numerical simulations. It became evident that plasma and laser wakefield acceleration studies are getting constantly more complex and demanding so that numerical tools are inevitable. While theory and analytical models remain necessary to clarify the relevant underlying physical mechanisms, numerics is the never ending help to support, drive and validate these theories. 

WG6 presentations could be subdivided into three main streams approaches: numerical simulations in support of physical problems or experimental campaigns, numerical simulations used to uncover new underlying mechanisms and the development of new numerical schemes in support of the previous goals. WG6 focused its attention on three main scientific topics: the mitigation of the numerical Cherenkov radiation, the mitigation of the hose-instability and on start-to-end simulations.

The theoretical approaches has focused on the understanding of the beam evolution and matching for a plasma wakefield acceleration case, with specific attention to the driver bunch for short and long running distances, and to exact symplectic approaches.

In summary the four sessions, and so the discussed topics, could be itemized as follow:
\begin{enumerate}
\item Analytical models
\begin{itemize}
\item Transverse bunch matching
\end{itemize}

\item Numerical Results
\begin{itemize}
\item Code development
\item Code update
\end{itemize}

\item Numerical Methods
\begin{itemize}
\item Numerical Cherenkov mitigation
\item Spectral methods PSATD
\item Adaptive Mesh Refinement (AMR)
\item Envelope
\end{itemize}

\item  Scientific computing
\begin{itemize}
\item Dynamic load balancing
\item Data Reduction
\item In situ visalization
\item Use of new libraries
\end{itemize}
\end{enumerate}

\section{Analytical models and numerical results}
The presented analytical models focused on the development of a closed form model to study the channel centroid in hosing instability for PWFA \cite{robson}, the approach considered leverage on a moments approach (very similar to the Boltzmann one, but in the form and fashion originally developed by Maxwell himself). A similar approach has also been used to study and predict the emittance evolution for an external injetion case \cite{aschikhin}. Along the same problem a model to predict the equilibrium condition for a driver bunch has been discusses in \cite{lotov}. The model consider that the head erosion plays a key role, but after several betatron oscillations, the driver bunch will eventually reach a stable shape and conformation, the theory is used to find this final configuration. The same author also highlights how numerical simulations generally are in need of a relatively short window simulation box, but one has to consider that the AWAKE experiment would need boxes that are three orders of magnitude longer \cite{lotov}, requiring new strategies and new approaches. Some theory has also been proposed for TNSA (target normal sheet acceleration), where with a simple electrostatic approach mixed with some initial conditions obtained by particle in cell code simulation, it is possible to calculate the maximum energy that can be reached by particles \cite{sinigardi}.

At the transition between theory and simulation, we find the evolution of the symplectic approach that allows for exact solution but (at present) still with computational constraints. Moving to more computationally oriented presentations, a possible strategy to mitigate the beam hosing instability with a betatron chirp in the quasi linear regime has been presented \cite{lehe}. External injection, with some start to end approach, loading particles from external tracking codes and remapping the background density where the bunches evolve from a MHD code, has also been presented \cite{marocchino}. Ionization has also played a key role in the presentations, and it has been considered as a valuable approach to produce in a controlled self-injected way a high brightness trailing bunch; this technique has been proposed both for the case of PWFA \cite{mira} as well as for LWFA \cite{lee}. LWFA new approaches, such as a bi-color stack of sub-joule pulses has been considered for Thomson scattering based gamma ray source \cite{Kalmykov}. The bi-color is resilient to degradation of the dense plasma and electron beam quality is preserved. 

\section{Numerical methods and implementations}

Numerical methods and their implementations are quickly improving in quality but also evolving in their implementations in order to follow the fast evolution of super-computers. 
Avoiding numerical Cherenkov effects, resulting from the unphysical interaction of relativistic macro-particles with their own field, is a priority
in accelerator simulations.
It appears that Pseudo-Spectral Analytical Time Domain (PSATD) solvers, virtually free of numerical dispersion, are not subject to numerical Cherenkov radiation while retaining satisying parallelization capabilities\cite{Jalas}.
Moreover, they can be used in a Galilean frame, co-moving with the plasma, in order to suppress numercial Cherenkov instability as well.
This allows simulations in a Lorentz boosted frame and significant speed-ups without compromising the accuracy of the results\cite{Kirchen}.
Another suggested way to improve accuracy is the numerical noise control technique. Controlled additional noise is introduced in a series of simulations in order to extrapolate the results to the very low noise level domain. This ultra low noise level is otherwise unreachable because it would require unreasonable numerical resolution\cite{Spitsyn}.

The WarpX project, targeting PIC simulations on exascale systems was presented\cite{Vay}. 
It relies on the already existing Warp code and the PICSAR library for the high performance computing (HPC) kernels thus combining very advanced algorithms.
It is also going to provide adaptative mesh refinement (AMR) by having overlapping grids of different resolutions.
Refined grids are framed by perfectly matched layers (PML) boundary conditions and driven by the surrounding coarser grid.

The latest advances in the PIConGPU code demonstrated that a very efficient cross-platform portability is achieved with a reasonable code size and complexity by using smart abstraction techniques and external libraries \cite{Huebl}. 
Advanced physics and diagnostics were also reported in the PIConGPU code \cite{Garten, Pausch}.
The ability to describe and take into account photon generation and their complex interaction with plasmas, including atomic physics, leads to a
brand new kind of simulations. 
This puts additional experiments within the reach of PIC simulations even if, in some cases, the PIC cycle represents only a fraction of the compute time.

\section{Code Comparison and Benchmarking}
Code comparison and code benchmarking is a common problem in the whole field and is still an \textit{open question}. Numerical codes are generally tested against theoretical solutions, however full theoretical solutions only exist for very simplified cases where numerical solutions themselves would not offer much more than pure mathematics. The interest is to use numerical codes where theory can, at present, hardly get. In \textit{classical numerics}, by solving simplified problems we can control the error that depends on the chosen numerical scheme and the resolution. In a \textit{multi-physics} scenario, as today codes are sometimes referred to, due to the variety of techniques used to integrate, interpolate and weighting, it is rather difficult to understand how far is the numerical solution from an ideal closed-form analytical solution. A possible way to tackle this issue is by comparing different codes with each other. Code comparison is a strategy that allows to verify that implementations are consistent and that base models are well aligned and coherent (theory is correct). Furthermore, understanding if all the relevant physical mechanisms have been taken into account requires comparisons with experiments. These might be rather difficult due to the large number of experimentally uncontrollable variables and physical mechanisms that concur during a single experiment. Designing a simple and clear experiment against which codes could be benchmarked might not be an easy task. Presentations during the workshop showed a clear motivation to work toward these goals. A clear example is the effort towards the \textit{synthetic diagnostics}: numerical codes are trying to reproduce laboratory diagnostics to shorten the distance between theory and laboratory experiments.

\section{Conclusions}

Theories and simulations of advanced accelerators is progressing quickly.
Software developpers are doing their best to keep up with most experimental ambitions but also with the constantly evolving super-computers.
Very few setups are now  out of reach of these tools but challenges of a different kind remains.
The first challenge is the increasing variety of techniques eventually leading to a lack of confidence in the different methods.
As explained in the previous section, thorough cross checking and benchmarking is now absolutely mandatory to guarantee the accuracy of the results.
Nevertheless, with the noticeable exception of openPMD, the lack of accepted standards makes this work terribly difficult.
The second challenge is the unprecedented numerical complexity which creates very high level entry barriers to the most efficient tools.
In this regard, collaboration between physicists and scientific computing experts is absolutely recommended.



\begin{thebibliography}{10}


\bibitem{robson}
R. Robson, T. Mehrling, J. Osterhoff
\newblock{NIMA} this conference

\bibitem{aschikhin}
A. Aschikhin et al.
\newblock{NIMA} this conference

\bibitem{lotov}
K. Lotov et al.
\newblock{NIMA} this conference

\bibitem{sinigardi}
S. Sinigardi et al.
\newblock{NIMA} this conference

\bibitem{lehe}
R. Lehe et al.
\newblock{NIMA} this conference

\bibitem{marocchino}
A. Marocchino
\newblock{NIMA} this conference

\bibitem{mira}
F. Mira
\newblock{NIMA} this conference

\bibitem{lee}
P. Lee
\newblock{NIMA} this conference


\bibitem{Kalmykov}
S. Kalmykov
\newblock{NIMA} this conference

\bibitem{Jalas}
S. Jalas
\newblock{NIMA} this conference

\bibitem{Kirchen}
M. Kirchen
\newblock{NIMA} this conference

\bibitem{Spitsyn}
R. Spitsyn
\newblock{NIMA} this conference

\bibitem{Vay}
J.-L. Vay
\newblock{NIMA} this conference

\bibitem{Huebl}
A. Huebl
\newblock{NIMA} this conference

\bibitem{Garten}
M. Garten
\newblock{NIMA} this conference

\bibitem{Pausch}
R. Pausch
\newblock{NIMA} this conference


\end{thebibliography}

%
%
\renewcommand{\baselinestretch}{1.0}

%
%

\end{document}